\documentstyle[seceq,mbf,preprint]{ptptex}
 \notypesetlogo 
 \def\ind{\indent}
 \def\nn{\nonumber}
 \def\be{\begin{equation}}
 \def\ee{\end{equation}}
 \def\hx{{\hat x}}
 \def\hS{{\hat S}}
 
 \def\htheta{{\hat \theta}}
 \def\hvarphi{{\hat \varphi}}
 \def\beq{\begin{eqnarray}}
 \def\eeq{\end{eqnarray}}
 \def\ba{\begin{array}}
 \def\ea{\end{array}}
 \def\cL{{\mbox {${\cal L}$}}}
 \def\cM{{\mbox {${\cal M}$}}}
 \def\cO{{\mbox {${\cal O}$}}}
 
 \def\bk{{\mbox {{\mbf k}}}}
  \def\bq{{\mbox {{\mbf q}}}}

 \def\bp{{\mbox {{\mbf p}}}}
 \def\cL{{\mbox {${\cal L}$}}}

\title{%
Lorentz Invariance And Unitarity Problem in \\
Non-Commutative Field Theory
}
\vspace{5mm}
\author{%
  Katsusada {\sc Morita},$^{1}$
  Yoshitaka {\sc Okumura}$^{2}$ and Eizou {\sc Umezawa}$^{3}$
}
\inst{%
{\it  $^{1}$Department of Physics, 
  Nagoya University, Nagoya 464-8602, Japan\\
  $^{2}$Department of Natural Sciences, 
  Chubu University, Kasugai 487-0027, Japan\\
  $^{3}$School of Health Sciences, Fujita Health University, 
  Toyoake 470-1192, Japan}
}
\recdate{%
\today
}
\vspace{5mm}
\abst{%
It is shown that the one-loop two-point amplitude 
in {\it Lorentz-invariant} non-commutative 
(NC) $\phi^3$ theory is finite after subtraction
in the commutative limit and satisfies the
usual cutting rule, thereby eliminating the unitarity problem 
in Lorentz-non-invariant NC field theory
in the approximation considered.
}
\begin{document}
\maketitle
\section{Introduction}
In the past few years many encouraging works have been done
on the non-commutative quantum field theory (NCQFT, simply called
QFT$_*$ in this paper),
that is, quantum field theory (QFT) based on the assumption 
that the space-time, say, at the Planck scale 
would become point-less with non-commuting coordinates
where gravity plays an essential role
and the usual notion of the space-time structure
may become ineffective.
The idea of non-commuting coordinates
dates back to Snyder\cite{1)} and Yang\cite{2)} in 1947
before the renormalization theory. But the idea
seems to have more profound meaning when 
gravity (black hole) and quantum mechanics (uncertainty
relation) is to be reconciled.\cite{3)}
It is a matter of course that
recent upsurge of QFT$_*$
mainly arises from a possible connection
of non-commutative geometry with string theory 
as refined by Seiberg and Witten.\cite{4)}
\\
\ind
QFT$_*$ is a field theory on the non-commutative space-time
in which the space-time coordinates represented by
hermitian operators $\hx^\mu$ no longer commute:
\be
[\hx^\mu,\hx^\nu]=i\theta^{\mu\nu},\quad
\quad\mu,\nu=0,1,2,3.
\label{eqn:1-1}
\ee
Here $(\theta^{\mu\nu})$ is a real antisymmetric 
constant matrix. 
We call (\ref{eqn:1-1}) the $\theta$-algebra.
Any field in QFT$_*$ is an operator-valued function, $\hvarphi(\hx)$.
Associated with it is the Weyl symbol
$\varphi(x)$ 
\begin{eqnarray}
\hvarphi(\hx)&=&
\int\!d^{\,4}k{\tilde\varphi}(k)e^{ik\hx}\longleftrightarrow
\varphi(x)=
\int\!d^{\,4}k{\tilde\varphi}(k)e^{ikx},
\label{eqn:1-2}
\end{eqnarray}
where $k\hx\equiv k_\mu\hx^\mu$. The operator product
$\hvarphi_1(\hx)\hvarphi_2(\hx)$ then corresponds to the
Moyal $\ast$-product of the Weyl symbols
\begin{eqnarray}
\hvarphi_1(\hx)\hvarphi_2(\hx)\longleftrightarrow
\varphi_1(x)*\varphi_2(x)\equiv
\varphi_1(x)e^{\frac i2
\theta^{\mu\nu}{\mbox{\scriptsize$
{\overleftarrow{{\partial_\mu}}}$}}
{\mbox{\scriptsize$\overrightarrow{\partial_\nu}$}}}
\varphi_2(x).
\label{eqn:1-3}
\end{eqnarray}
The Weyl-Moyal correspondence implies that QFT$_*$ is a nonlocal
field theory
defined on the ordinary space-time with
the point-wise multiplication of field variables
being replaced by the $\ast$-product. Thus 
the action defining QFT$_*$ is given by
\be
S={\rm tr}[{\hat{\cL}}(\hvarphi(\hx), \partial_\mu\hvarphi(\hx))]
=\int\!d^{\,4}x\cL(\varphi(x), \partial_\mu\varphi(x))_*,
\label{eqn:1-4}
\ee
where we have normalized tr$e^{ik\hx}=(2\pi)^4\delta^4(k)$
and the subscript of the Lagrangian indicates that
the $\ast$-product should be taken for all products
of the field variables. 
\\
\ind
This nonlocality makes the theory difficult to manage
in comparison with QFT although it may resolve at least partially
the divergence problem.\cite{5),6)}
The main source of the difficulty
is due to the so-called IR/UV mixing\cite{7)}:
the nonplanar Feynman diagram is made UV finite,
but instead shows IR singularity when the regularization
parameter is removed.
Another source of the difficulty
concerns with a possible breakdown\cite{8)} of the
familiar theorems established in QFT.
One notable example is the unitarity violation.\cite{9)}
An approach based on the
Hamiltonian formalism
to avoid the unitarity problem
was proposed in Ref. 10). It was elaborated by Rim and
Yee.\cite{11)}
Chu et al.\cite{12)}
discussed the interplay between the IR/UV
mixing and the unitarity problem.
\\
\ind
Focusing on the unitarity problem,
the present paper proposes another approach.
It grows from our careful repeating of
the calculation by Gomis and Mehen\cite{9)} of the one-loop
amplitude for the two-point function
in the NC $\phi^3$ ($\phi^*{}^3$) model.
We found that, if the Feynman amplitude is Lorentz-invariant,
their proof of the unitarity relation
for the time-like external momentum
suggests non-appearance of the
imaginary part in the space-like region
on the contrary to what claimed by them,
thereby eliminating
the unitarity problem in the approximation under consideration.
\\
\ind
The next problem is how to
make the NC amplitude Lorentz-invariant.
In other words, it is necessary to construct
a Lorentz-invariant NC field theory.
Snyder\cite{1)} was the first who showed that
Lorentz invariance allows in addition to the continuum
space-time a NC space-time in which a single fundamental length
is naturally introduced. 
Doplicher, Fredenhagen and Roberts (DFR)\cite{3)}
also proposed a Lorentz-invariant NC field theory
whose Feynman rules are formulated by Filk.\cite{5)}
The DFR quantum space\cite{3)} is based on the algebra
\beq
[{\hat x}^\mu,{\hat x}^\nu]&=&i{\hat\theta}^{\mu\nu},\;\;\;
[{\hat\theta}^{\mu\nu},{\hat x}^\nu]=0=
[{\hat\theta}^{\mu\nu},{\hat\theta}^{\rho\sigma}],
\quad\mu,\nu,\rho,\sigma=0,1,2,3.
\label{eqn:1-5}
\eeq
where $\htheta^{\mu\nu}$ is an antisymmetric second-rank
tensor operator.
It was rediscovered by Carlson, Carone and Zobin\cite{13)} 
who introduced the Lorentz-invariant action
described below.
We have called (\ref{eqn:1-5}) the DFR algebra
in Refs. 14) and 15). As shown in Ref. 15)
the $\theta$-algebra (\ref{eqn:1-1})
holds true as a `weak' relation,
where $\theta^{\mu\nu}$ is an eigenvalue
of the operator $\htheta^{\mu\nu}$.
Thus $\theta^{\mu\nu}$ is an antisymmetric second-rank
tensor. This permits us to define a fundamental length $a$
through\cite{14)}
\beq
\theta^{\mu\nu}&=&a^2{\bar\theta}^{\mu\nu}.
\label{eqn:1-6}
\eeq
\ind
Based on this algebra
the connections (\ref{eqn:1-2})
and (\ref{eqn:1-3}) are maintained
by a slight modification.
Namely, the Weyl symbol now depends on the eigenvalue $\theta^{\mu\nu}$ of 
the operator $\htheta^{\mu\nu}$, written as $\varphi(x,\theta)$,
and one needs an integration over the extra 6-dimensional
variable $\theta^{\mu\nu}$.
The new correspondence is
\begin{eqnarray}
&&\hvarphi(\hx,\htheta)=
\int\!d^{\,4}kd^{\,6}\sigma
{\tilde\varphi}(k,\sigma)e^{ik\hx+i\sigma\htheta}\longleftrightarrow
\varphi(x,\theta)=
\int\!d^{\,4}kd^{\,6}\sigma{\tilde\varphi}(k,\sigma)e^{ikx+i\sigma\theta}\nn\\[2mm]
&&\hvarphi_1(\hx,\htheta)\hvarphi_2(\hx,\htheta)\longleftrightarrow
\varphi_1(x,\theta)*\varphi_2(x,\theta)
\equiv
\varphi_1(x,\theta)e^{\frac i2
\theta^{\mu\nu}{\mbox{\scriptsize$
{\overleftarrow{{\partial_\mu}}}$}}
{\mbox{\scriptsize$\overrightarrow{\partial_\nu}$}}}
\varphi_2(x,\theta),
\label{eqn:1-7}
\end{eqnarray}
where $\sigma\htheta=\sigma_{\mu\nu}\htheta^{\mu\nu}$.
The action (\ref{eqn:1-4})
is then replaced with
\be
\hS={\rm tr}[{\hat{\cL}}(\hvarphi(\hx,\htheta), 
\partial_\mu\hvarphi(\hx,\htheta))]
=\int\!d^{\,4}xd^{\,6}\theta\,
W(\theta)\cL(\varphi(x,\theta), \partial_\mu\varphi(x,\theta))_*.
\label{eqn:1-8}
\ee
This form of the Lorentz-invariant QFT$_*$ action with a
normalized weight function $W(\theta)$ being Lorentz-invariant
was first obtained by Carlson, Carone and Zobin,\cite{13)}
applied to construct Lorentz-invariant NCQED
\cite{14)}
and further studied in Ref.15).
\\
\indent
In the framework of the Lorentz-invariant
NC field theory
we reinvestigated 
the unitarity problem.
Taking $D=4$ $\phi^*{}^3$ theory in the one-loop
approximation for the self-energy diagram, 
we carefully repeat in the next section
Gomis-Mehen's calculation
and found that the unitarity relation
even for time-like momentum
is frame-dependent in their theory.
Facing this problem
we then ask ourselves
what happens on the unitarity problem
if we employ the Lorentz-invariant NC field theory
with the action (\ref{eqn:1-8}).
We obtain in the section 3 an interesting result
that the one-loop amplitude
for the self-energy diagram
in the Lorentz-invariant $\phi^*{}^3$ theory
is finite after subtraction in the
commutative limit and unitary.
We give some remarks in the last section.
\section{A comment on the paper by Gomis and Mehen} 
For our purpose, it is enough to consider one of the models
discussed in Ref. 9), namely, $D=4$
$\phi^*{}^{3}$ theory based on (\ref{eqn:1-1}).
The action is given by
\beq
S&=&\int\!d^4x[\frac 12\partial_\mu\phi(x)*
\partial^\mu\phi(x)-\frac 12m^2\phi(x)*\phi(x)
-\frac {\lambda}{3!}\phi(x)*
\phi(x)*\phi(x)],
\label{eqn:2-1}
\eeq
where $\phi(x)$ is a scalar field, $m$ is 
the mass parameter and $\lambda$ is the
coupling constant.
The amplitude $M_{ab}$ for the transition $a\to b$
must satisfy the unitarity
relation,
\beq
2{\rm Im}\,M_{ab}&=&
\sum_nM_{an}M_{bn}^*,
\label{eqn:2-2}
\eeq
provided $M_{ab}=M_{ba}$ which is true in the following.
The sum is to be taken over all possible intermediate states.
\\
\ind
In this section, following Gomis and Mehen,\cite{9)}
we confine ourselves to check the unitarity relation
to the order $\cO(\lambda^2)$.
The one-loop amplitude for the self-energy diagram
is given by
\begin{eqnarray}
M
&=&-i\frac{\lambda^2}4\int\!\frac{d^4q}{(2\pi)^4}
\frac{1+\cos{(p\wedge q)}}{((p-q)^2-m^2+i\epsilon)(q^2-m^2+i\epsilon)},
\label{eqn:2-3}
\end{eqnarray}
where $p$ is the external momentum, 
and $p\wedge q=p_\mu\theta^{\mu\nu}q_\nu$.
Using Feynman parameter and Schwinger representation
we write it as a convergent integral
\beq
M&=&\frac{\lambda^2}{64\pi^2}
\int_0^1\!dx\int_0^\infty\!d\alpha \alpha^{-1}
\big(e^{-\alpha(-x(1-x)p^2+m^2-i\epsilon)}+
e^{-\alpha(-x(1-x)p^2+m^2-i\epsilon)-\frac {p\circ p}{4\alpha}}\big)
e^{-\frac 1{4\alpha \Lambda^2}},
\label{eqn:2-4}
\eeq
where
we define
\beq
p\circ p&\equiv&p^\mu\theta^2_{\mu\nu}p^\nu
=(p^0\theta_{0i})^2-(p^i\theta_{i0})^2
-2p^0\theta_{0i}\theta_{ij}p^j
+(p_i\theta^{ij})^2,
\label{eqn:2-5}
\eeq
and the last factor $e^{-\frac 1{4\alpha \Lambda^2}}$
is the regularization one except for which (\ref{eqn:2-4})
is identical to (the analytical continuation of)
Eq.(2.8) in Ref. 9). 
\footnote{We have evaluated the amplitude
(\ref{eqn:2-3}) via Wick rotation 
and then analytically continued \'a la Gomis and Mehen.
The convergence of the integral (\ref{eqn:2-4})
limits $p^2<4m^2$ and $p\circ p+\frac 1{\Lambda^2}>0$.}
It is important to remember that
$M$ depends on $p\circ p$ as well as $p^2,
M=M(p^2, p\circ p)$, where the regularization dependence
is not displayed.
If $\theta^{\mu\nu}$ is a tensor,
$p\circ p$ is Lorentz-invariant.
This is not the case if $\theta^{\mu\nu}$ is a constant
matrix as usually assumed. The Lorentz violation in the conventional
NC field theory manifests itself in the fact that
$p\circ p$ is {\it not} Lorentz-invariant for constant $\theta^{\mu\nu}$,
neither is the amplitude $M$.
\\
\ind
It is straightforward to obtain from (\ref{eqn:2-4})
\beq
M&=&\frac{\lambda^2}{32\pi^2}\int_0^1\!dx
[K_0\big(\sqrt{(-x(1-x)p^2+m^2)/\Lambda^2}\big)\nn\\[2mm]
&&\qquad\qquad
+K_0\big(\sqrt{(p\circ p+1/\Lambda^2)(-x(1-x)p^2+m^2)}\big)],
\label{eqn:2-6}
\eeq
where $K_0$ is the modified Bessel function.
This gives the imaginary part for $p^2>4m^2$
\beq
{\rm Im}\,M&=&\frac{\lambda^2}{64\pi}\int_{\frac{1-\gamma}2}
^{\frac{1+\gamma}2}\!dx
[J_0\big(\sqrt{(p^2x(1-x)-m^2)/\Lambda^2}\big)\nn\\[2mm]
&&\qquad\qquad
+J_0\big(\sqrt{(p\circ p+1/\Lambda^2)(p^2x(1-x)-m^2)}\big)],
\label{eqn:2-7}
\eeq
where $\gamma=\sqrt{1-\frac{4m^2}{p^2}}$, $J_0$ is the Bessel function
(put $n=0$ in (\ref{eqn:3-26})).
Since the imaginary part is regularization-independent,
we let $\Lambda^2\to \infty$ to recover Gomis-Mehen's result\cite{9)}
\beq
{\rm Im}\,M&=&\frac{\lambda^2}{64\pi}\gamma
+\frac{\lambda^2}{64\pi}\int_{\frac{1-\gamma}2}
^{\frac{1+\gamma}2}\!dxJ_0\big(\sqrt{p\circ p(p^2x(1-x)-m^2)}\big)\nn\\[2mm]
&=&\frac{\lambda^2}{64\pi}\gamma
+\frac{\lambda^2}{32\pi}
\frac{\sin{(\gamma\sqrt{p^2p\circ p}/2)}}{\sqrt{p^2p\circ p}},
\label{eqn:2-8}
\eeq
where we have assumed $p\circ p>0$. Recall that
this imaginary part originates from the divergence at 
$\alpha\to \infty$
of the $\alpha$-integral in (\ref{eqn:2-4}) for $p^2>4m^2$.
\\
\ind
Let us now go on to the evaluation of the unitarity sum (\ref{eqn:2-2})
derived from the cutting rule
in the same approximation.
Gomis and Mehen\cite{9)} presented the following result
for $p^2>4m^2$
\beq
\sum|M|^2&=&
\frac {\lambda^2}{2(2\pi)^2}
\int\!\frac{d^3{\mbf k}}{2\omega_{\mbf k}}
\int\!\frac{d^3{\mbf q}}{2\omega_{\mbf q}}
\delta^4(p-q-k)\frac {1+\cos{(p\wedge k)}}2\nn\\[2mm]
&=&
\frac{\lambda^2}{4\cdot 32\pi^2}\gamma
\int\!d\Omega[1+\cos{(p\wedge k)}]
=\frac{\lambda^2}{32\pi}\gamma
+\frac{\lambda^2}{16\pi}
\frac{\sin{(\gamma\sqrt{p^2p\circ p}/2)}}{\sqrt{p^2p\circ p}}.
\label{eqn:2-9}
\eeq
Note that there are two on-shell particles in the intermediate state,
their momenta being labelled by $k$ and $q=p-k$.
This result shows that the unitarity relation
(\ref{eqn:2-2}) for $a=b$ being the one-particle
state with $p^2>4m^2$ for which
$p\circ p>0$
\footnote{This is so because
(\ref{eqn:2-5}) implies $p\circ p=(p^0\theta_{0i})^2$
in the rest frame, $p^0\ne 0, \bp=0$, for time-like case.
If $p\circ p$ is {\it not} Lorentz scalar,
it may take different numerical values in different Lorentz frames,
but the unitarity sum should be taken in the same Lorentz frame
where the imaginary part is evaluated.}
is satisfied to order $\cO(\lambda^2)$.
For space-like momentum for which $p\circ p$ can become negative,
we are unable to let $\Lambda^2\to \infty$ 
keeping the positivity of $p\circ p+\frac 1{\Lambda^2}$.
This signals the breakdown of the unitarity relation,
for the unitarity sum identically vanishes for space-like momentum
and the imaginary part of (\ref{eqn:2-6})
when $p\circ p+\frac 1{\Lambda^2}$ is negative is given by
\beq
{\rm Im}\,M&=&\frac{\lambda^2}{64\pi}\int_0^1
\!dx
J_0\big(\sqrt{|p\circ p+1/\Lambda^2|(-x(1-x)p^2+m^2)}\big)\nn\\[2mm]
&=&\frac{\lambda^2}{64\pi}\int_0^1
\!dx
J_0\big(\sqrt{|p\circ p|(-x(1-x)p^2+m^2)}\big),\;\;\;(p^2<0),
\label{eqn:2-10}
\eeq
where we take the limit $\Lambda^2\to \infty$.
This imaginary part
comes from the divergence of the $\alpha$-integral
at $\alpha\to 0$ (not $\infty$) 
in the second term in (\ref{eqn:2-4}) due to the presence
of the extra term $-\frac{p\circ p}{4\alpha}$ in the exponential.
This is the critical observation by Gomis and Mehen.\cite{9)}
The next problem we focus on in this section is to carefully
compute the integral appearing in (\ref{eqn:2-9}),
\beq
I&=&\int\!d\Omega\cos{(p\wedge k)}.
\label{eqn:2-11}
\eeq 
If $p\circ p$ is not Lorentz scalar,
its value differs in general in different
Lorentz frames. If, instead, $p\circ p$ is Lorentz scalar,
its value in a particular Lorentz frame
is the same for all Lorentz frames.
In what follows we compute the integral
(\ref{eqn:2-11}) in two different Lorentz frames.
\\
\ind
For $p$ time-like we go over to
the rest frame, $p^0\ne 0, \bp=0$ so that $p\circ p=(p^0\theta_{0i})^2>0$.
In the rest frame we have
\beq
p\wedge k&=&p_0\theta^{0i}k_i
={\tilde {\mbf p}}\cdot\bk=|{\tilde {\mbf p}}||\bk|
\cos{\theta},
\label{eqn:2-12}
\eeq
where $\theta$ is the angle between $\bk$ and
\beq
{\tilde {\mbf p}}=
({\tilde p}_1=\theta^{01}p_0,
{\tilde p}_2=\theta^{02}p_0,
{\tilde p}_3=\theta^{03}p_0)
\label{eqn:2-13}
\eeq
It follows from
(\ref{eqn:2-5}) that
\beq
p\circ p&=&|{\tilde{\mbf p}}|^2
\label{eqn:2-14}
\eeq
in the rest frame. Consequently, we have
\beq
p\wedge k&=&\sqrt{p\circ p}|\bk_{\rm cm}|\cos{\theta}
\label{eqn:2-15}
\eeq
where we have used the fact that
$|\bk|$ in the rest frame
is given by the invariant center of mass momentum
$|\bk_{\rm cm}|=\sqrt{\frac{p^2}4-m^2}$.
Thus we arrive at the result
\beq
I&=&2\pi\int_0^\pi\!d\theta\sin{\theta}\cos{(\sqrt{p\circ p}|\bk_{\rm cm}|\cos{\theta})}
=4\pi
\frac{\sin{(\sqrt{p\circ p}|\bk_{\rm cm}|)}}{\sqrt{p\circ p}|\bk_{\rm cm}|},
\label{eqn:2-16}
\eeq
hence obtaining (\ref{eqn:2-9}). Note that
we have only checked the unitarity relation
in a particular Lorentz frame, the rest frame for time-like $p$.
The same value of $p\circ p$
as given by (\ref{eqn:2-14})
has to be substituted into the imaginary part, (\ref{eqn:2-8}).
\\
\ind
There exists another Lorentz frame in which $\theta^{0i}=0$,
where the integral $I$ is easily calculated.
\footnote{This never means that
the unitarity sum leads to a non-trivial result also in this case.}
In this frame, 
we have $p\circ p=(p_i\theta^{ij})^2$ and
\beq
p\wedge k&=&
p_i\theta^{ij}k_j
={\tilde{\mbf p}'}\cdot\bk
=|{\tilde{\mbf p}'}||\bk|\cos{\theta'}
=\sqrt{p\circ p}|\bk|\cos{\theta'}\nn\\[2mm]
{\tilde {\mbf p}'}&=&
({\tilde p}'_1=\theta^{j1}p_j,
{\tilde p}'_2=\theta^{j2}p_j, {\tilde p}'_3=\theta^{j3}p_j)\nn\\[2mm]
p\circ p&=&|{\tilde{\mbf p}'}|^2.
\label{eqn:2-17}
\eeq
Thus (\ref{eqn:2-16}) is regained if we put $|\bk|=|\bk_{\rm cm}|$.
It should be remarked, however, that,
although we used the same notation $p\circ p$
in (\ref{eqn:2-15}) and (\ref{eqn:2-17}),
they are defined
in a different way in the two cases
so that they may be numerically different. In this sense
the unitarity relation in NC field theory is frame-dependent.
In order to check it we should evaluate both sides of the unitarity relation
in the same Lorentz frame, but
the resulting relation holds true separately
in the Lorentz frames employed.
\\
\ind
If the amplitude (\ref{eqn:2-3}) is considered to be Lorentz-invariant,
the situation is drastically changed.
In such a case, the unitarity relation is also Lorentz-invariant,
and it should be true in any Lorentz frames if it is proved in a
particular Lorentz frame. However,
the Lorentz frames $\theta^{0i}=0$
and $\theta^{ij}=0$ are not connected by any Lorentz
transformations just like
time-like $p$ is never Lorentz-transformed into
space-like $p$.
\footnote{They are only connected through
analytic continuation.}
If the Lorentz frame $\theta^{0i}=0$
is connected with the rest frame by Lorentz transformations,
both $|{\tilde{\mbf p}}|^2$ and $|{\tilde{\mbf p}'}|^2$ 
should be the same numerically.
If, on the other hand, the Lorentz frame $\theta^{ij}=0$
is connected with the rest frame by Lorentz transformations,
both $|{\tilde{\mbf p}}|^2$ and $(p^0\theta_{0i})^2-(p^i\theta_{i0})^2$
\footnote{This is always positive for time-like $p$
because, by Schwarz inequality ,
$(p^0\theta_{0i})^2-(p^i\theta_{i0})^2\ge
(p^0\theta_{0i})^2-(p^i)^2(\theta_{i0})^2
=p^2(\theta_{i0})^2>0$.}
should be the same numerically.
In any case, it is enough to integrate (\ref{eqn:2-12})
in the rest frame by assuming that
$p\circ p$ is Lorentz scalar.
\\
\ind
In the next section we argue that
all Lorentz frames for time-like $p$ 
have negative value of $\zeta$,
\beq
\zeta=\frac 12\theta^{\mu\nu}\theta_{\mu\nu},
\label{eqn:2-18}
\eeq
while any Lorentz frames for space-like $p$
have positive value of $\zeta$.
There exists a Lorentz frame $\theta^{ij}=0$ for $\zeta<0$
and for $\zeta>0$ we can choose a Lorentz frame with $\theta^{0i}=0$.
Hence, $p\circ p=(p^0\theta_{0i})^2$
for time-like $p$ in the rest frame, while $p\circ p=(p_i\theta^{ij})^2$
for space-like $p$
in the Lorentz frame with $\theta^{0i}=0$.
Consequently, $p\circ p$
is positive definite for both time-like and
space-like momenta.
That is,
Lorentz invariance cures the unitarity problem.
The imaginary part
(\ref{eqn:2-10}) completely disappears from the scene because
$p\circ p+1/\Lambda^2$ is positive even
for space-like $p$.
\\
\ind
For this reason
we expect that
Lorentz invariance is a powerful guiding
principle also in NC field theory.
However, it is impossible to regard
$\theta^{\mu\nu}$ in (\ref{eqn:1-1}) as an antisymmetric
$c$-number tensor if $\hx^\mu$ transforms as a 4-vector.\cite{15)}
What this means is that we can not regard the amplitude (\ref{eqn:2-3})
as being Lorentz-invariant.
Consequently, we must look for another NC algebra
which naturally provides us with a tensor $\theta^{\mu\nu}$,
thus enabling us to
prove our conjecture alluded to above with a bit certainty,
and preserves Lorentz invariance of the theory.
Such a Lorentz-invariant NC field theory connected with the $\theta$-algebra
was formulated by
Carlson, Carone and Zobin\cite{13)}
based on the DFR algebra (\ref{eqn:1-5}).
In the next section we prove the absence of the unitarity
problem at one-loop in the Lorentz-invariant 
$\phi^*{}^3$ theory.
\section{Lorentz invariance and unitarity problem in
$\phi^*{}^{3}$ theory} 
Let us now consider the same model based on the DFR algebra (\ref{eqn:1-5}).
The Lorentz-invariant action is given by
\beq
\hS&=&\int\!d^{\,6}\theta\, W(\theta)
\int\!d^{\,4}x[\frac 12\partial_\mu\phi(x)*
\partial^\mu\phi(x)-\frac 12m^2\phi(x)*\phi(x)
-\frac {\lambda}{3!}\phi(x)*
\phi(x)*\phi(x)],
\label{eqn:3-1}
\eeq
where the scalar field $\phi(x)$ is assumed to be
independent of $\theta$. Using (\ref{eqn:1-6}) we put
\beq
W(\theta)&=&a^{-12}w({\bar\theta}).
\label{eqn:3-2}
\eeq
We may then rewrite (\ref{eqn:3-1}) as
\beq
\hS&=&\int\!d^{\,6}{\bar\theta}\, w({\bar\theta})
\int\!d^{\,4}x[\frac 12\partial_\mu\phi(x)*
\partial^\mu\phi(x)-\frac 12m^2\phi(x)*\phi(x)
-\frac {\lambda}{3!}\phi(x)*
\phi(x)*\phi(x)].
\label{eqn:3-3}
\eeq
The vertex in the Feynman diagram is associated with
$i\lambda$ times the factor
\beq
V(k_1,k_2)&=&\int\!d^{\,6}{\bar\theta}\, w({\bar\theta})
\cos{\frac{k_1\wedge k_2}2},
\label{eqn:3-4}
\eeq
where $k_1$ and $k_2$ are the momenta flowing into
the vertex. In the previous section we
did not encounter the ${\bar\theta}$-integration.
\\
\ind
Using the above Feynman rule we obtain
the one-loop amplitude for the self-energy diagram
\beq
\cM
&=&-i\frac{\lambda^2}2\int\!\frac{d^{\,4}q}{(2\pi)^4}
\frac{V^2(p,q)}{((p-q)^2-m^2+i\epsilon)(q^2-m^2+i\epsilon)}.
\label{eqn:3-5}
\end{eqnarray}
In order to evaluate this integral
we first determine the vertex factor $V(p,q)$.
As a model we take the one
discussed in Ref. 14).
Using the formulae (4$\cdot$ 14) in Ref. 14)
we find 
\beq
V(p,q)&=&\int\!d^{\,6}{\bar\theta}\, w({\bar\theta})
[1-\frac 1{2!}(\frac{p\wedge q}2)^2
+\frac 1{4!}(\frac{p\wedge q}2)^4-+\cdots]\nn\\[2mm]
&=&1-\frac{a^4}{2!4}\frac{\langle{\bar\theta}^{\,2}\rangle}{12}
[q^2p^2-(p\cdot q)^2]
+\frac{a^8}{4!4^2}\frac{\langle{\bar\theta}^{\,4}\rangle}{64}
[q^2p^2-(p\cdot q)^2]^2-+\cdots\nn\\[2mm]
&=&
e^{-\frac{a^4}4\frac{\langle{\bar\theta}^{\,2}\rangle}{24}[q^2p^2-(p\cdot q)^2]},
\label{eqn:3-6}
\end{eqnarray}
where the invariant moments are defined by
\begin{eqnarray}
\langle{\bar\theta}^{\,2n}
\rangle=\int\!d^{\,6}{\bar\theta}\,w({\bar\theta})
({\bar\theta}^{\mu\nu}{\bar\theta}_{\mu\nu})^n, n=0,1,2,\cdots,
\label{eqn:3-7}
\end{eqnarray}
and we exponentiated the infinite sum by assuming the relations
\beq
\langle{\bar\theta}^{\,4}\rangle&=&
\frac 43(\langle{\bar\theta}^{\,2}\rangle)^2
\label{eqn:3-8}
\eeq
and so on. These relations are obtained for Gaussian
$w({\bar\theta})$.\cite{14),15)}
\\
\ind
For time-like $p$ we choose the rest frame, $p_0\ne 0, \bp=0$
in which
\beq
V(p,q)&=&e^{\frac{a^4}4\frac{\langle{\bar\theta}^{\,2}\rangle}{24}p_0^2{\mbf q}^2}
\equiv J.
\label{eqn:3-9}
\eeq
To ensure the convergence of the integral (\ref{eqn:3-5})
we have to assume that
\beq
\langle{\bar\theta}^{\,2}\rangle<0\;\;\;{\rm for}\;
{\rm time{\mbox{-}}like}\;p.
\label{eqn:3-10}
\eeq
We then compute the integral (\ref{eqn:3-5})
by first enclosing the $q_0$-integration path from
$-\infty$ to $\infty$ 
with the upper large semi-circle in the $q_0$-plane
in a counterclockwise way. This picks up two
poles at $q_0=-\omega_{\mbf q}+i\epsilon$
and at $q_0=p_0-\omega_{\mbf p-q}+i\epsilon
=p_0-\omega_{\mbf q}+i\epsilon$,
where $\omega_{\mbf q}=\sqrt{\bq+m^2}$.
The result turns out to be
\beq
\cM&=&-\frac{\lambda^2}{2(2\pi)^3}
\int\!d^3\bq\frac{J^2}{(2\omega_{\mbf q})^2}
\big[\frac 1{p_0-2\omega_{\mbf q}+i\epsilon}
-\frac 1{p_0+2\omega_{\mbf q}+i\epsilon}\big].
\label{eqn:3-11}
\eeq
Thanks to the assumption (\ref{eqn:3-10})
this integral is convergent.
For $p_0>0$ this gives the imaginary part
\beq
{\rm Im}\,\cM&=&
\frac{\lambda^2}{2(2\pi)^3}\int\!d^3\bq\frac{J^2}{(2\omega_{\mbf q})^2}
\pi\delta(p_0-2\omega_{\mbf q})\nn\\[2mm]
&=&
\frac{\lambda^2}{8\pi}\frac{|{\mbf q}|J^2}{2p_0}
=\frac{\lambda^2}{32\pi}\gamma e^{
\frac {a^4}2\frac{\langle{\bar\theta}^2\rangle}{24}(p^2)^2(\frac \gamma2)^2},
\label{eqn:3-12}
\eeq
where $\gamma$ is the same as in (\ref{eqn:2-7}).
The final result is an invariant one valid for
$p^2>4m^2$.
This is nothing but half the unitarity sum
\beq
\sum|\cM|^2&=&
\frac {\lambda^2}{2(2\pi)^2}
\int\!\frac{d^3{\mbf k}}{2\omega_{\mbf k}}
\int\!\frac{d^3{\mbf q}}{2\omega_{\mbf q}}
\delta^4(p-q-k)J^2\nn\\[2mm]
&=&\frac{\lambda^2}{16\pi}\gamma e^{
\frac {a^4}2\frac{\langle{\bar\theta}^2\rangle}{24}(p^2)^2(\frac \gamma2)^2}.
\label{eqn:3-13}
\eeq
That is, we proved the unitarity relation
\beq
2{\rm Im}\,\cM&=&\sum|\cM|^2,
\label{eqn:3-14}
\eeq
for the two-point function
in the lowest-order approximation.
\\
\ind
Now that we have checked the unitarity relation
to order $\cO(\lambda^2)$ in a Lorentz-invariant way,
we proceed to prove that the Lorentz-invariant
amplitude (\ref{eqn:3-5}) do not develop the imaginary part
for space-like $p$. 
For space-like $p$ we choose the Euclidean momenta
\beq
p^0&\to& ip_E^4,\;\;\;
\bp\to \bp_E,\;\;\;p^2\to -p_E^2,\nn\\[2mm]
q^0&\to& iq_E^4,\;\;\;
\bq\to \bq_E,\;\;\;q^2\to -q_E^2,
\label{eqn:3-15}
\eeq
where the Euclidean momenta $p_E$ and $q_E$
are real.
In addition the non-commutativity parameter
is also made Euclidean,
\beq
\theta^{0i}\to -i\theta_E^{4i},\;\;\;
\theta^{ij}\to \theta_E^{ij}
\label{eqn:3-16}
\eeq
so that
\beq
p\wedge q=p_E\wedge_Eq_E\equiv
\sum_{\mu,\nu=1,2,3,4}(p_E)_\mu\theta_E^{\mu\nu}(q_E)_\nu.
\label{eqn:3-17}
\eeq
The vertex factor $V(p,q)$ becomes the {\it invariant
damping factor}\cite{14)}
\beq
V(p_E,q_E)&=&
e^{-\frac{a^4}4\frac{\langle{\bar\theta}^{\,2}_E\rangle}{24}[q_E^2p_E^2-(p_E\cdot q_E)^2]},
\label{eqn:3-18}
\end{eqnarray}
since
\beq
\langle{\bar\theta}^{\,2}\rangle
\to\langle{\bar\theta}_E^{\,2}\rangle>0\;\;\;{\rm for}\;
{\rm space{\mbox{-}}like}\;p.
\label{eqn:3-19}
\eeq
The amplitude (\ref{eqn:3-5})
becomes for
space-like $p$
\beq
\cM
&=&\frac{\lambda^2}{32\pi^4}
\int_0^1\!dx\int\!d^{\,4}q_E
\frac{V^2(p_E,q_E)}{(q_E^2+\mu^2_E)^2},
\label{eqn:3-20}
\end{eqnarray}
where we put $\mu^2_E=p_E^2x(1-x)+m^2$.
Using Schwinger representation
we rewrite this expression as
\beq
\cM
&=&\frac{\lambda^2}{32\pi^4}
\int_0^1\!dx\int_0^\infty\! d\alpha \alpha
\int\!d^{\,4}q_E
e^{-\alpha(q_E^2+\mu^2_E)-A_E[q_E^2p_E^2-(p_E\cdot q_E)^2]},
\label{eqn:3-21}
\end{eqnarray}
with $A_E=\frac {a^4}2\frac{\langle {\bar\theta}^{\,2}_E\rangle}{24}>0$.
The $q_E$-integration is easily done
in the frame $p_E^1=p_E^2=0$ to obtain an invariant result
\beq
\cM
&=&\frac{\lambda^2}{32\pi^2}
\int_0^1\!dx\int_0^\infty\! d\alpha 
\frac{\sqrt{\alpha}}{(\sqrt{\alpha+A_Ep_E^2})^3}
e^{-\alpha\mu^2_E}.
\label{eqn:3-22}
\eeq
If we go back to the Minkowski space,
we have
\beq
\cM
&=&\frac{\lambda^2}{32\pi^2}
\int_0^1\!dx\int_0^\infty\! d\alpha 
\frac{\sqrt{\alpha}}{(\sqrt{\alpha-Ap^2})^3}
e^{-\alpha\mu^2},
\label{eqn:3-23}
\eeq
where $A=\frac {a^4}2\frac{\langle {\bar\theta}^{\,2}\rangle}{24}$
is positive for space-like $p$
since it was positive in the Euclidean metric
\footnote{Although $\zeta$ of (\ref{eqn:2-18})
is indefinite
in the Lorentz metric,
$A$ remains positive
in the Minkowski space for space-like $p$.
This is understood by remarking that
$p_E^2>0$
in the Euclidean metric goes over to $-p^2>0$
in the Lorentz metric, while space-like $p$
is always space-like in any Lorentz frames.
The analytic continuation, say,
$p^2\to e^{it}p^2$ and
$\langle {\bar\theta}^2\rangle\to
e^{-it}\langle {\bar\theta}^2\rangle$
with $t=0\to \pi$ (or $-\pi$) converts space-like $p$
into time-like $p$ accompanied with
the sign change of $\langle {\bar\theta}^2\rangle$.
It is so chosen that the function $(\sqrt{\alpha-Ap^2})^3$
is single-valued.}
and $\mu^2=-p^2x(1-x)+m^2$ which is $\ge m^2$ for space-like $p$.
There are two facts to be noted.
First, since $-Ap^2>0$ for space-like $p$,
the amplitude $\cM$ has no imaginary
part in that region. This is due to the fact that
the imaginary part for space-like $p$,
if exist, comes from a possible divergence
at $\alpha\to 0$ of the $\alpha$-integral in (\ref{eqn:3-23}),
which, in fact, converges for space-like $p$, $-Ap^2>0$.
In other words, there is no unitarity
problem in the one-loop level for 1-1 transition in the
Lorentz-invariant $\phi^*{}^3$ model.
Second,
the amplitude $\cM$ includes both planar and nonplanar
diagrams and the integral (\ref{eqn:3-23})
is finite as far as $-Ap^2=
-\frac {a^4}2\frac{\langle {\bar\theta}^2\rangle}{24}
p^2>0$.
A possible divergence at $p^2\to 0$
is connected with the log divergence in the
commutative limit. Namely,
the IR limit $p^2\to 0$
is indistinguishable from the 
commutative limit $a\to 0$, where
we should recover the well-known
log divergence which is subtracted off by renormalization.
This is the correspondence principle
that QFT$_*$ should satisfy.
Although we are far from formulating it in a quantitative way,
we may define the regularized amplitude by
\beq
\!\!\!\!\!\!\!\!\!
\cM_{\rm reg}(p^2)&=&\cM(p^2)-\cM(0)
=\frac{\lambda^2}{32\pi^2}
\int_0^1\!dx\int_0^\infty\! d\alpha 
\big[\frac{\sqrt{\alpha}}{(\sqrt{\alpha-Ap^2})^3}
e^{-\alpha\mu^2}-\frac{\sqrt{\alpha}}{(\sqrt{\alpha})^3}
e^{-\alpha m^2}\big].
\label{eqn:3-24}
\eeq
The integrand in (\ref{eqn:3-24})
at $\alpha\to 0$ when $a=0$ behaves like
$\frac{\sqrt{\alpha}}{(\sqrt{\alpha})^3}
[e^{\alpha p^2x(1-x)}-1]
e^{-\alpha m^2}\to \frac1\alpha[\alpha p^2x(1-x)]
e^{-\alpha m^2}$
which renders the integral (\ref{eqn:3-24})
convergent at $\alpha\to 0$.
\\
\ind
It would be interesting to directly 
check the unitarity relation
from (\ref{eqn:3-23}).
To this purpose we need to remember (\ref{eqn:3-10})
so that $-Ap^2$ is also positive for time-like $p$.
That is, we require that $A$ changes sign
if $p^2$ is analytically continued from negative to
positive values to make the function
$(\sqrt{\alpha-Ap^2})^3$ single-valued.
If otherwise, the calculation leading to
(\ref{eqn:3-12}) will not be justified.
The resulting imaginary part
stems from the divergence of the integral
(\ref{eqn:3-23}) at $\alpha \to \infty$.
Moreover, we expand the factor
$\frac{\sqrt{\alpha}}{(\sqrt{\alpha-Ap^2})^3}$
with respect to $-Ap^2$, which introduces
additional divergences at $\alpha \to 0$. To avoid
these new divergences
we multiply the integrand of (\ref{eqn:3-23}) 
through the regularization factor $e^{-\frac 1{\alpha\Lambda^2}}$.
Consequently, we must evaluate the integral
for $0<p^2<4m^2$ where $\mu^2>0$ for $0\le x\le 1$
\beq
\cM
&=&\frac{\lambda^2}{32\pi^2}
\int_0^1\!dx\int_0^\infty\! d\alpha 
\frac{\sqrt{\alpha}}{(\sqrt{\alpha-Ap^2})^3}
e^{-\alpha\mu^2-\frac 1{\alpha\Lambda^2}}\nn\\[2mm]
&=&\frac{\lambda^2}{32\pi^2}
\int_0^1\!dx
\sum_{n=0}^\infty
\left(
\ba{c}
-\frac 32\\
n\\
\ea
\right)
(-Ap^2)^n\int_0^\infty\! \frac{d\alpha}{\alpha^{n+1}}
e^{-\alpha\mu^2-\frac 1{\alpha\Lambda^2}}\nn\\[2mm]
&=&
\frac{\lambda^2}{32\pi^2}
\int_0^1\!dx
\sum_{n=0}^\infty
\left(
\ba{c}
-\frac 32\\
n\\
\ea
\right)
(-Ap^2)^n2(\frac 1{\Lambda^2\mu^2})^{-\frac n2}
K_{-n}(2\sqrt{\mu^2/\Lambda^2}).
\label{eqn:3-25}
\eeq
Since the imaginary part of the modified Bessel function
with the argument $\mu^2=e^{-i\pi}|\mu^2|$
is given by
\beq
{\rm Im}\,
\big[2(\frac 1{\Lambda^2|\mu^2|})^{-\frac n2}
e^{-n\pi i/2}
K_{-n}(2e^{-i\pi/2}\sqrt{|\mu^2|/\Lambda^2})\big]
&=&\pi (\frac 1{\Lambda^2|\mu^2|})^{-\frac n2}
J_n(2\sqrt{|\mu^2|/\Lambda^2})
\label{eqn:3-26}
\eeq
and the imaginary part for $p^2>4m^2$ 
is independent of the regularization parameter $\Lambda^2$,
we let $\Lambda^2\to \infty$ with $J_n(x)\to (\frac x2)^n\frac 1{n!}$
at $x\to 0$ to obtain for $p^2>4m^2$
\beq
{\rm Im}\,\cM&=&\frac{\lambda^2}{32\pi^2}
\int_{\frac{1-\gamma}2}^{\frac{1+\gamma}2}\!dx
\sum_{n=0}^\infty
\left(
\ba{c}
-\frac 32\\
n\\
\ea
\right)
(-Ap^2)^n\pi\frac{|\mu^2|^n}{n!}\nn\\[2mm]
&=&
\frac{\lambda^2}{32\pi}
\gamma\sum_{n=0}^\infty\frac 1{n!}(A(p^2)^2)^n
(\frac{\gamma^2}4)^n
=
\frac{\lambda^2}{32\pi}\gamma
e^{A(p^2)^2(\frac \gamma2)^2}.
\label{eqn:3-27}
\eeq
This
is identical with the previous result (\ref{eqn:3-12}).
\section{Remarks} 
Unitarity originates from the probability interpretation
of quantum mechanics and, hence, is 
one of the indispensable elements in any theory
subject to quantum mechanical interpretation.
It is formulated as an asymptotic completeness in QFT.
The asymptotic states are classified according to
the unitary irreducible representations
of the Poincar\'e group which
must be the symmetry group of the underlying
space-time. The quantum space\cite{3)} based on the
DFR algebra enjoys the Poincar\'e symmetry.
However, the
symmetry group of the $\theta$-algebra
is smaller than the Poincar\'e group.
In the canonical form of the
non-commutativity parameter
the $\theta$-algebra has\cite{8)} the symmetry
$O(1,1)\times SO(2)\bowtie T_4$, where
$\bowtie$ denotes the semi-direct product.
This implies, for instance, that
the tachyonic states to be
excluded from
the asymptotic states by the spectral condition
in QFT
may be classified into `massive' states
according to the symmetry group
$O(1,1)\times SO(2)\bowtie T_4$,
which appear in the intermediate states in a closure relation.\cite{8)}
These unsatisfactory aspects are expected to
disappear if we `relativitize' the $\theta$-algebra
in favor of the DFR algebra.
\\
\ind
Our proof of the absence of the unitarity problem
in the Lorentz-invariant NC field theory
was made in the lowest-order approximation
in a simplest unrealistic model. Nonetheless, we expect that
unitarity is valid 
in more general circumstance.
\\
\ind
We also found that
the amplitude (\ref{eqn:3-23})
is finite as far as $-Ap^2>0$.
The subtraction is made by noting that
the IR limit $p^2\to 0$ can not be distinguished
from the commutative limit $a\to 0$ where
we should recover the well-known log divergence
if QFT$_*$ satisfies the correspondence principle.
\footnote{Long wave length `sees' the space-time
in a coarse way, that is,
in the IR limit, the space-time non-commutativity
loses its meaning. An interpretation along this direction
was put forward in Ref. 14).}
The problem of quantifying the
correspondence principle
is, however, left over.
\section*{Acknowledgements}
The authors are grateful to
H. Kase for useful discussions and
careful reading of the manuscript.

\end{document}